\documentclass[a4paper]{article}
\usepackage{color}
\usepackage{romannum}
\usepackage{INTERSPEECH_v2}
\usepackage{cleveref}
\title{What Does the Speaker Embedding Encode?}
\name{Shuai Wang, Yanmin Qian, Kai Yu\thanks{This work was supported by the Shanghai Sailing Program No. 16YF1405300, the China NSFC projects (No. 61573241 and No. 61603252) and the Interdisciplinary Program (14JCZ03) of Shanghai Jiao Tong University in China. Experiments have been carried out on the PI supercomputer at Shanghai Jiao Tong University.}}
\address{
     Key Lab. of Shanghai Education Commission for Intelligent Interaction and Cognitive Engineering \\
    SpeechLab, Department of Computer Science and Engineering \\
    Brain Science and  Technology Research Center \\
    Shanghai Jiao Tong University, Shanghai, China
    }
\email{wsstriving@gmail.com, yanminqian@sjtu.edu.cn, kai.yu@sjtu.edu.cn}

\begin{document}

\maketitle

\begin{abstract} 
Developing a good speaker embedding has received tremendous interest in the speech community, with representations such as {\it i}-vector and {\it d}-vector demonstrating remarkable performance across various tasks. Despite their widespread adoption, a fundamental question remains largely unexplored: what properties are actually encoded in these embeddings? To address this gap, we conduct a comprehensive analysis of three prominent speaker embedding methods: {\it i}-vector, {\it d}-vector, and RNN/LSTM-based sequence-vector ({\it s}-vector). Through carefully designed classification tasks, we systematically investigate their encoding capabilities across multiple dimensions, including speaker identity, gender, speaking rate, text content, word order, and channel information. Our analysis reveals distinct strengths and limitations of each embedding type: {\it i}-vector excels at speaker discrimination but encodes limited sequential information; {\it s}-vector captures text content and word order effectively but struggles with speaker identity; {\it d}-vector shows balanced performance but loses sequential information through averaging. Based on these insights, we propose a novel multi-task learning framework that integrates {\it i}-vector and {\it s}-vector, resulting in a new speaker embedding ({\it i-s}-vector) that combines their complementary advantages. Experimental results on RSR2015 demonstrate that the proposed {\it i-s}-vector achieves more than 50\% EER reduction compared to the {\it i}-vector baseline on content mismatch trials, validating the effectiveness of our approach.
\end{abstract}


\section{Introduction}
Speaker recognition, the task of identifying individuals through their voice characteristics, has witnessed remarkable progress over the past decades. Depending on whether the spoken text is constrained, speaker recognition systems can be categorized as text-dependent or text-independent. Both paradigms have been extensively studied, leading to significant performance improvements in recent years.

The evolution of speaker recognition has been marked by several key milestones. Since its introduction \cite{reynolds2000speaker}, the Gaussian Mixture Model-Universal Background Model with maximum a posteriori adaptation (GMM-UBM-MAP) framework dominated the field for many years. In this approach, speaker models are derived from a universal background model (UBM) through MAP adaptation. However, GMM-UBM-MAP suffers from data sparsity issues during enrollment, where only a subset of UBM parameters are effectively adapted \cite{reynolds2000speaker, shum2011unsupervised}. To address this limitation, researchers explored methods to correlate different Gaussian components through global transformations, leading to the super-vector representation \cite{campbell2006support, campbell2006svm}, which achieved competitive results when combined with support vector machines.

The pursuit of fixed-length vector representations gained momentum with the introduction of Joint Factor Analysis (JFA) \cite{kenny2005joint}, which modeled speaker and channel factors in separate subspaces within the super-vector space. This was subsequently simplified by the {\it i}-vector approach \cite{dehak2011front}, which models a single total variability subspace, becoming the de facto standard for speaker recognition.

The success of deep learning in speech recognition \cite{hinton2012deep} catalyzed its adoption in speaker recognition. Early efforts integrated DNNs into the {\it i}-vector framework by replacing GMMs with DNN-based acoustic models \cite{lei2014novel,kenny2014deep}. Alternative approaches emerged, including DNN-based bottleneck feature extraction \cite{fu2014tandem,liu2015deep, tian2015investigation,richardson2015unified} and direct speaker representation learning \cite{variani2014deep,li2015improved,chen2015multi,tian2015investigation}. Among these, {\it d}-vector \cite{variani2014deep} has become a prominent representative. More recently, researchers have explored utterance-level representations using recurrent neural networks, particularly long short-term memory (LSTM) networks \cite{heigold2016end,bhattacharyadeep,hochreiter1997long}.

Despite the proliferation of speaker embedding methods and their demonstrated effectiveness \cite{dehak2011front,variani2014deep,chen2015multi,heigold2016end,bhattacharyadeep}, a fundamental question remains largely unanswered: \textit{what properties are actually encoded in these embeddings?} Understanding the encoding characteristics of different embeddings is crucial for selecting appropriate methods for specific applications and for developing improved representations. Inspired by fine-grained analysis methodologies for sentence embeddings in natural language processing \cite{adi2016fine}, we propose a systematic investigation framework to analyze speaker embeddings. Through carefully designed classification tasks, we reveal the following key insights:

\leftmargini=1mm
\begin{itemize}
    \item {\it i}-vector demonstrates superior speaker discrimination capability, making it highly effective for speaker identity tasks. While it can encode spoken terms to a considerable extent, it fails to capture sequential information such as word order.
    \item LSTM/RNN-based sequence-vector ({\it s}-vector) excels at encoding text content and word order information, but is less effective at modeling speaker identity compared to {\it i}-vector.
    \item {\it d}-vector shows balanced performance across multiple properties but loses sequential information due to its averaging operation.
    \item A carefully designed combination of {\it i}-vector and {\it s}-vector can leverage their complementary strengths, resulting in a more powerful speaker embedding.
\end{itemize}

The main contributions of this work are threefold: (1) we propose a systematic framework for analyzing what properties are encoded in different speaker embeddings; (2) we provide comprehensive empirical analysis comparing {\it i}-vector, {\it d}-vector, and {\it s}-vector across multiple dimensions; (3) we introduce a novel multi-task learning architecture that integrates {\it i}-vector and {\it s}-vector, achieving significant performance improvements on text-dependent speaker verification tasks.

The remainder of this paper is organized as follows. Section 2 introduces the three speaker embedding methods under investigation. Section 3 describes our analysis methodology. Section 4 presents experimental results and detailed analysis. Section 5 introduces our proposed {\it i-s}-vector framework and presents results on the RSR2015 dataset. Section 6 concludes the paper.

\section{Speaker Embedding Methods}

This section provides an overview of the three speaker embedding methods analyzed in this work: {\it i}-vector, {\it d}-vector, and {\it s}-vector.

\subsection{{\it i}-vector}
The {\it i}-vector framework \cite{dehak2011front} represents a speaker- and session-dependent super-vector $\mathbf{M}$ (derived from a UBM) as a linear transformation in a low-dimensional subspace:
\begin{equation}
\mathbf{M}=\mathbf{m}+\mathbf{Tw}
\end{equation}
where $\mathbf{m}$ is a speaker- and session-independent super-vector (typically the UBM mean super-vector), $\mathbf{T}$ is a low-rank total variability matrix that captures both speaker and session variability, and $\mathbf{w}$ is a latent vector. The {\it i}-vector is defined as the posterior mean of $\mathbf{w}$ given the observed acoustic features. This compact representation (typically 400-600 dimensions) has become the standard in speaker recognition due to its effectiveness and robustness.

\subsection{{\it d}-vector}
The {\it d}-vector approach \cite{variani2014deep} leverages deep neural networks to learn speaker-discriminative representations. A DNN is first trained in a supervised manner to classify speakers, using frame-level acoustic features as input. After training, frame-level vectors are extracted from the last hidden layer of the DNN. The utterance-level {\it d}-vector is obtained by averaging these frame-level representations over the entire utterance. This averaging operation, while providing a fixed-length representation, inherently discards temporal ordering information. The model architecture for {\it d}-vector extraction is illustrated in \Cref{fig:dvector}.

\subsection{{\it s}-vector}
\begin{figure}[!tbp]
  \centering
  \begin{minipage}[b]{0.23\textwidth}
    \includegraphics[width=\textwidth]{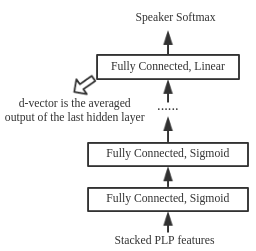}
    \caption{{\it d}-vector extraction}
    \label{fig:dvector}
  \end{minipage}
  \hfill
  \begin{minipage}[b]{0.23\textwidth}
    \includegraphics[width=\textwidth]{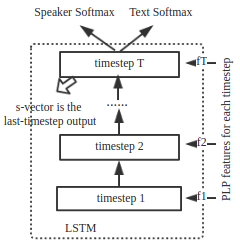}
    \caption{{\it s}-vector extraction}
    \label{fig:svector}
  \end{minipage}
\end{figure}

The {\it s}-vector (sequence-vector) approach treats speaker recognition as a sequence modeling problem, leveraging recurrent neural networks, particularly LSTM networks, to capture temporal dependencies. As discussed in \cite{bhattacharyadeep}, sequence-level representations can be obtained through various strategies: last-timestep embedding, attention-based embedding, or averaged embedding. In this work, we employ the last-timestep approach, where the hidden state at the final timestep serves as the sequence representation ({\it s}-vector). This approach naturally preserves sequential information, making it well-suited for capturing word order and temporal patterns. However, training LSTM models for speaker recognition can be challenging due to limited utterance-level training samples. To address this, we adopt a multi-task learning framework similar to \cite{chen2015multi}, where the model is trained to predict both speaker identity and text content simultaneously. The model architecture is shown in \Cref{fig:svector}.

\section{Analysis Methodology}

Our analysis methodology is grounded on a fundamental principle: \textbf{if a property is encoded in a speaker embedding, it should be possible to train a classifier to predict that property from the embedding, and the classification accuracy reflects the extent to which the property is encoded}. This principle enables us to systematically investigate what information different embeddings capture.

Given speaker embeddings extracted from utterances, we construct classification tasks to probe specific properties. We focus on three broad categories of properties:

\begin{itemize}
\item \textbf{Speaker-related properties}: speaker identity, gender, and speaking rate.
\item \textbf{Text-related properties}: spoken terms (word presence), word order, and utterance length.
\item \textbf{Channel-related properties}: handset identity and recording channel characteristics.
\end{itemize}

We analyze three speaker embedding types: {\it i}-vector, {\it d}-vector, and {\it s}-vector. For each embedding type and each property, we extract embeddings, construct labeled datasets, and train classifiers to predict the target property. To ensure that classification performance reflects the embedding's inherent encoding capability rather than classifier sophistication, we employ a simple Multi-Layer Perceptron (MLP) with a single hidden layer and ReLU activation for all experiments. This design choice allows us to attribute performance differences to the embeddings themselves rather than to classifier complexity.

\subsection{Prediction Tasks}
Based on the methodology described above, we design eight prediction tasks to probe different properties encoded in speaker embeddings:

\leftmargini=1mm
\begin{itemize}

    \item  \textbf{Speaker Identity Task}: This task evaluates the extent to which speaker embeddings encode speaker identity, which is the core objective of speaker recognition. The evaluation set contains 106 different speakers, forming a 106-class classification problem.

    \item  \textbf{Speech Text Task}: This task assesses whether speaker embeddings capture information about the spoken text content. The dataset contains 30 different sentences, resulting in a 30-class sentence classification task.
    
    \item  \textbf{Spoken Term Task}: This task investigates whether embeddings encode information about which specific words are present in an utterance. The dataset contains 147 distinct words. For each word, we train a binary classifier (logistic regression) to predict whether that word appears in the utterance. An embedding is considered correctly classified if all words in the corresponding utterance are correctly identified.
    
    \item \textbf{Word Order Task}: This task evaluates the ability of embeddings to capture sequential information, specifically word order. We formulate this as a ``chunk order" prediction task: given two utterances $\mathbf{u}_1$ and $\mathbf{u}_2$, and their concatenation $\mathbf{s}$, the goal is to predict whether $\mathbf{u}_1$ appears before or after $\mathbf{u}_2$ in $\mathbf{s}$, based on their speaker embeddings $\mathbf{se}_{u1}$, $\mathbf{se}_{u2}$, and $\mathbf{se}_{s}$. This is modeled as a binary classification task, where the input is the concatenation of the three embeddings. Positive and negative samples are generated by flipping the order of $\mathbf{u}_1$ and $\mathbf{u}_2$ in $\mathbf{s}$.
    
    \item \textbf{Utterance Length Task}: This task measures whether embeddings encode information about utterance duration. Based on utterance length (after voice activity detection), we categorize samples into four classes: 1-3s, 4-6s, 7-9s, and 10-12s (with gaps preserved for better discrimination). Longer utterances are created by concatenating shorter ones. A four-class classifier is trained for this task.
        
    \item \textbf{Channel Task}: This task investigates whether embeddings capture channel-related information, which can be a source of unwanted variability. The dataset contains recordings from 6 different handsets, corresponding to 6 distinct channels. A six-class classifier is trained to predict the recording channel.

    \item  \textbf{Speaker Gender Task}: This task evaluates whether embeddings encode speaker gender information. A binary classifier is trained to predict gender from the speaker embedding.

    \item  \textbf{Speaking Rate Task}: This task assesses whether embeddings capture speaking rate information. The original dataset contains only normal-speed speech. We generate additional samples by time-stretching utterances to $0.5 \times$ and $1.5 \times$ the original speed, creating three classes: slow, normal, and fast. A three-class classifier is trained for this task.

\end{itemize}

\section{Experiments and Analysis}

\subsection{Experimental Setup}
All experiments are conducted on the RSR2015 part 1 dataset \cite{larcher2014text}, which consists of 300 speakers (143 females and 157 males). Each speaker pronounces 30 fixed sentences selected from the TIMIT database \cite{garofolo1993darpa}, with 9 recording sessions per sentence, ensuring coverage of all English phonemes. The average recording duration is 3.20 seconds. The dataset is divided into three subsets: background ({\it bkg}), development ({\it dev}), and evaluation ({\it eval}). 

\begin{table}[!htbp]
\caption{Subset definition of RSR2015 part 1}\label{tab:rsr2015}
\centering
\begin{tabular}{| c |c | c | c |}
  \hline
  Subset & \# Female speaker& \# Male speaker & \# Total \\
  \hline
  \hline
  bkg & 47 & 50 & 97\\
  dev & 47 & 50 & 97\\ 
  eval & 49 & 57 & 106\\
  \hline
\end{tabular}

\end{table}

The {\it bkg} subset is used for training all embedding extractors: GMM-UBM and {\it i}-vector extractor, {\it d}-vector DNN extractor, and {\it s}-vector LSTM extractor. All experiments use 39-dimensional PLP features (13-dimensional static features with delta and delta-delta coefficients). The {\it eval} subset is used exclusively for the prediction tasks described in Section 3.

The implementation details for each embedding method are as follows:

\begin{itemize}

    \item  \textbf{{\it i}-vector}: 
    The {\it i}-vector system consists of a 1024-component GMM-UBM and {\it i}-vector extractors with varying dimensions. The entire pipeline is implemented using Kaldi \cite{povey2011kaldi}.

    \item \textbf{{\it d}-vector}:
    The DNN architecture consists of: (1) an input layer with 429 neurons (39-dimensional features with left and right context frames), (2) 5 hidden layers with 1024 neurons each, (3) an extraction layer with $|vector|$ neurons (where $|vector|$ is the desired embedding dimension), and (4) an output softmax layer with 97 neurons (one for each speaker in the {\it bkg} set). Sigmoid activation is used throughout. The network is initialized using Restricted Boltzmann Machines (RBMs) and then fine-tuned using stochastic gradient descent (SGD) with backpropagation. After training, the {\it d}-vector is obtained by averaging frame-level representations extracted from the last hidden layer over the entire utterance.

    \item  \textbf{{\it s}-vector}:
    The LSTM architecture initially consisted of a 39-dimensional input layer, a standard LSTM hidden layer, and a softmax output layer with 97 neurons (one per speaker). By varying the number of LSTM units in the hidden layer, we can extract {\it s}-vectors of different dimensions. However, training LSTM models utterance-wise results in limited training samples, leading to poor generalization. To address this, we employ a multi-task learning approach similar to \cite{chen2015multi}, where the model has two output layers: one for speaker classification (97 neurons) and one for text classification (30 neurons). This multi-task framework provides additional supervision signals and improves model training. The LSTM is optimized using the RMSprop algorithm \cite{tieleman2012lecture}.

\end{itemize}

In all experimental figures, the green line labeled {\it i-s}-vector (based on unidirectional LSTM) represents the newly proposed speaker embedding, which will be described in Section 5.

\subsection{Speaker Identity Task}
\begin{figure}[htb]
  \centering
  \includegraphics[width=0.45\textwidth]{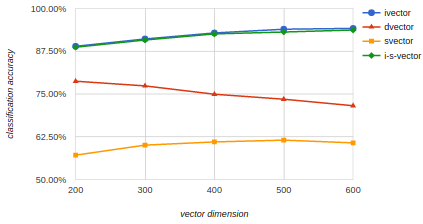}
  \caption{Speaker identity task}
  \label{fig:spktask}
\end{figure}

Speaker discrimination capability is fundamental to speaker recognition systems. As shown in Figure \ref{fig:spktask}, {\it i}-vector significantly outperforms the other two embedding methods, achieving the highest classification accuracy. In contrast, {\it s}-vector performs worst on this task. This performance gap can be attributed to the fact that LSTM/RNN models operate at the utterance level, resulting in fewer training samples compared to frame-level approaches, which negatively impacts their generalization capability for speaker discrimination. 

\subsection{Speech Text Task}

\begin{figure}[htb]
  \centering
  \includegraphics[width=0.42\textwidth]{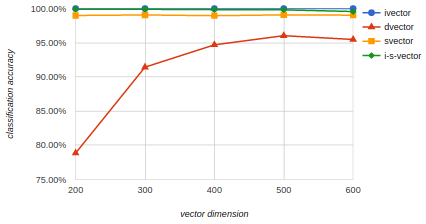}
  \caption{Speech text task}
  \label{fig:texttask}
\end{figure}

As shown in Figure \ref{fig:texttask}, both {\it i}-vector and {\it s}-vector achieve excellent performance on speech text classification, with both approaching 100\% accuracy. Interestingly, {\it d}-vector also achieves nearly 95\% accuracy despite averaging frame-level representations, which might be expected to lose text information. To better understand this observation and investigate the underlying mechanisms, we conduct additional analysis through the \textit{spoken term task}, \textit{word order task}, and \textit{utterance length task}. 

\subsubsection{Spoken Term Task}
\begin{figure}[!htb]
  \centering
  \includegraphics[width=0.42\textwidth]{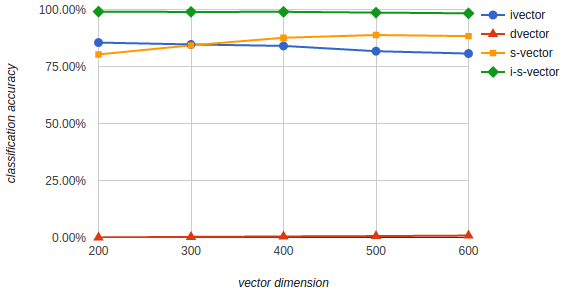}
  \caption{Speech content task}
  \label{fig:binarytask}
\end{figure}

This task evaluates whether speaker embeddings encode information about which specific words are present in an utterance. As shown in Figure \ref{fig:binarytask}, {\it s}-vector outperforms {\it i}-vector when the embedding dimension exceeds 300, demonstrating its superior capability in capturing word-level information. In contrast, {\it d}-vector fails to encode word information and cannot reliably predict spoken terms. This is consistent with expectations, as the averaging operation in {\it d}-vector extraction removes most word-level discriminative information. 

\subsubsection{Word Order Task}
\begin{figure}[htb]
  \centering
  \includegraphics[width=0.42\textwidth]{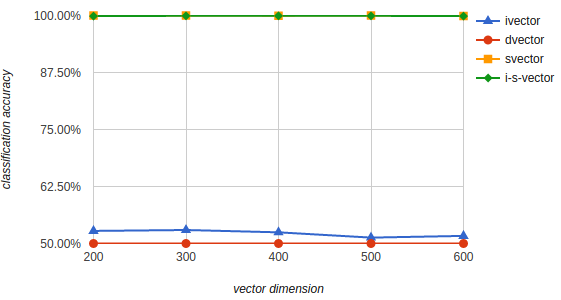}
  \caption{Word order task}
  \label{fig:ordertask}
\end{figure}

Figure \ref{fig:ordertask} compares the performance of different speaker embeddings on the word order task. Since this is a binary classification task, the random baseline is 50\%. As expected, {\it d}-vector fails to preserve any order information, performing at baseline level. {\it i}-vector also performs poorly on this task, confirming its inability to capture sequential information. In contrast, {\it s}-vector demonstrates superior performance, achieving nearly 100\% accuracy. These results align with the architectural characteristics of each method: {\it i}-vector and {\it d}-vector do not explicitly model temporal order, while recurrent networks inherently capture sequential dependencies through their recurrent structure.

\subsubsection{Utterance Length Task}
\begin{figure}[!htb]
  \centering
  \includegraphics[width=0.42\textwidth]{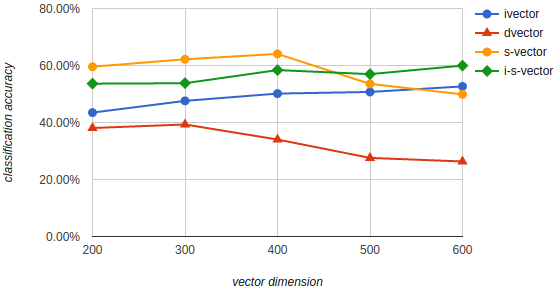}
  \caption{Length task}
  \label{fig:lengthtask}
\end{figure}

This task evaluates whether embeddings encode information about utterance duration. With four length categories, the random baseline accuracy is 25\%. As shown in Figure \ref{fig:lengthtask}, both {\it s}-vector and {\it i}-vector outperform {\it d}-vector, as expected. This suggests that these methods retain some temporal duration information, while {\it d}-vector's averaging operation largely removes this information. 

\subsection{Channel Task}
\begin{figure}[!htb]
  \centering
  \includegraphics[width=0.42\textwidth]{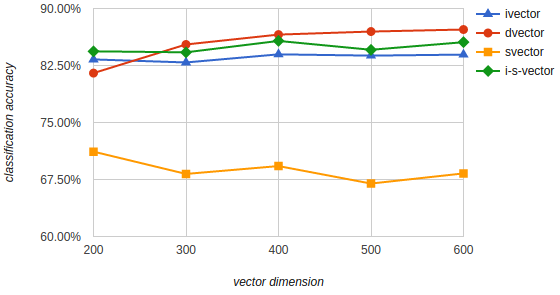}
  \caption{Channel task}
  \label{fig:channeltask}
\end{figure}

The RSR2015 dataset contains recordings from 6 different handsets, corresponding to 6 distinct channels. As shown in Figure \ref{fig:channeltask}, all three embedding types achieve prediction accuracies significantly higher than the random baseline (16.7\%). This indicates that these embeddings inadvertently encode channel-related information, which can be problematic as it makes them sensitive to channel mismatch and recording conditions. This finding highlights the importance of channel compensation techniques when deploying these embeddings in practical applications.

\subsection{Speaker Gender and Speaking Rate Tasks}
For the speaker gender task, all three embedding types achieve high accuracy (above 97\%), with {\it d}-vector approaching 100\%. This suggests that gender information is strongly encoded in all embedding types, likely due to fundamental acoustic differences between male and female voices.

Regarding speaking rate prediction, {\it i}-vector, {\it d}-vector, and {\it s}-vector achieve accuracies of 90\%, 95\%, and 77\% respectively. {\it d}-vector performs best on this task, possibly because frame-level DNN representations capture temporal dynamics more effectively than utterance-level aggregations.

\section{Combining {\it i}-vector and {\it s}-vector}

Our analysis reveals that different speaker embeddings exhibit complementary strengths: {\it i}-vector excels at speaker discrimination, while {\it s}-vector is superior at encoding text content and sequential information. For text-dependent speaker verification (TDSV), both speaker discriminative and text information are crucial. This motivates us to develop a unified framework that combines the advantages of both approaches.

\begin{figure}[!htb]
  \centering
  \includegraphics[width=0.3\textwidth]{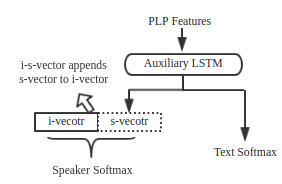}
  \caption{Proposed {\it i-s}-vector framework for TDSV}
  \label{fig:combinetask}
\end{figure}

We propose a multi-task learning framework that integrates {\it i}-vector and {\it s}-vector into a unified representation, termed {\it i-s}-vector. The key innovation is to train an LSTM model that incorporates both representations during training. Specifically, we concatenate the last-timestep hidden output ({\it s}-vector) with the corresponding {\it i}-vector before feeding the combined representation to the speaker classification softmax layer. The entire architecture is optimized end-to-end using multi-task learning, where the model simultaneously predicts speaker identity and text content. This approach differs from simple concatenation by allowing the network to learn optimal feature interactions during training. The architecture is illustrated in Figure \ref{fig:combinetask}.

As demonstrated in \Cref{fig:spktask,fig:texttask,fig:binarytask,fig:ordertask,fig:lengthtask}, the proposed {\it i-s}-vector achieves competitive or superior performance across almost all analysis tasks, demonstrating its ability to effectively combine the complementary strengths of {\it i}-vector and {\it s}-vector.

\subsection{Evaluation on RSR2015 Part 1}

We evaluate the proposed {\it i-s}-vector on the RSR2015 part 1 dataset for text-dependent speaker verification. Three test conditions are defined based on different impostor scenarios: (\Romannum{1}) content mismatch (correct speaker, wrong text), (\Romannum{2}) speaker mismatch (wrong speaker, correct text), and (\Romannum{3}) both mismatch (wrong speaker, wrong text). Results are reported in Table \ref{tab:eer} using cosine similarity as the scoring metric.

\begin{table}[!htb]
\caption{EER(\%) comparison of TDSV on RSR2015 part1}
\centering
\begin{tabular}{ c c c c }
\hline
\textbf{vectors/conditions}    & \textbf{\Romannum{1}} & \textbf{\Romannum{2}}  & \textbf{\Romannum{3}}  \\ \hline
\textit{i}-vector   & 0.35   &  \textbf{1.13}   & 0.06    \\ \hline
\textit{i}-vector + \textit{s}-vector   & 0.28   &  \textbf{1.13}   & 0.03   \\ \hline
\textit{i-s}-vector (LSTM) & 0.17  & 1.98 &0.03      \\ 
\textit{i-s}-vector (BLSTM) &\textbf{ 0.11} & 1.72 & \textbf{0.02 }      \\ \hline
\end{tabular}
\label{tab:eer}
\end{table}

The first row shows the {\it i}-vector baseline, which achieves performance comparable to state-of-the-art results reported in the literature \cite{larcher2014text,dey2016deep}. The second row shows results from direct concatenation of {\it i}-vector and {\it s}-vector. The bottom two rows show results from the proposed {\it i-s}-vector framework using unidirectional and bidirectional LSTM, respectively.

Key observations: (1) The proposed {\it i-s}-vector achieves more than 50\% EER reduction compared to the {\it i}-vector baseline on content-mismatch conditions (\Romannum{1} and \Romannum{3}), demonstrating its effectiveness in capturing text-dependent information. (2) Using bidirectional LSTM (BLSTM) provides further improvements, leveraging both forward and backward temporal context. (3) There is a performance degradation in condition \Romannum{2} (speaker mismatch with correct text), which suggests that the increased focus on text information may slightly reduce pure speaker discrimination capability. However, this trade-off is beneficial for text-dependent scenarios where content matching is crucial. (4) The proposed multi-task learning framework outperforms simple concatenation, validating the importance of joint optimization. These results confirm that {\it i-s}-vector is particularly effective for ``speaker-dependent content matching problems'' \cite{dey2016deep}, which are central to text-dependent speaker verification.

\section{Conclusion}

This paper presents a comprehensive analysis of three prominent speaker embedding methods: {\it i}-vector, {\it d}-vector, and LSTM/RNN-based sequence-vector ({\it s}-vector). Through systematically designed classification tasks, we investigate what properties are encoded in each embedding type. Our analysis reveals several key findings:

\leftmargini=3mm
\begin{itemize}

    \item {\it i}-vector demonstrates superior speaker discrimination capability, making it highly effective for speaker identity tasks. The GMM-UBM phoneme modeling enables {\it i}-vector to encode speech content information, but it fails to capture word order due to its bag-of-features nature.
    
    \item LSTM-based {\it s}-vector excels at encoding speech content and word order information, making it particularly valuable for tasks requiring sequential understanding, such as random digit speaker recognition.
    
    \item {\it d}-vector shows balanced performance across multiple properties but loses sequential information through its averaging operation, limiting its effectiveness for order-sensitive tasks.
    
    \item All embedding types inadvertently encode channel information to some extent, highlighting the importance of channel compensation techniques in practical deployments.
    
    \item The complementary strengths of {\it i}-vector and {\it s}-vector can be effectively combined through a carefully designed multi-task learning framework. The proposed {\it i-s}-vector achieves more than 50\% EER reduction compared to the {\it i}-vector baseline on content mismatch trials of RSR2015 part 1, demonstrating the practical value of understanding and leveraging embedding characteristics.
\end{itemize}

Our work provides insights that can guide the selection and design of speaker embeddings for specific applications, and demonstrates how understanding embedding characteristics can lead to improved representations through principled combination strategies.

\bibliographystyle{IEEEtran}

\bibliography{mybib}

@article{adi2016fine,
  title={Fine-grained analysis of sentence embeddings using auxiliary prediction tasks},
  author={Adi, Yossi and Kermany, Einat and Belinkov, Yonatan and Lavi, Ofer and Goldberg, Yoav},
  journal={arXiv preprint arXiv:1608.04207},
  year={2016}
}

@article{bhattacharyadeep,
  title={Deep Neural Networks based Text-Dependent Speaker Verification},
  author={Bhattacharya, Gautam and Alam, Jahangir and Stafylakis, Themos and Kenny, Patrick}
}

@article{reynolds2000speaker,
  title={Speaker verification using adapted Gaussian mixture models},
  author={Reynolds, Douglas A and Quatieri, Thomas F and Dunn, Robert B},
  journal={Digital signal processing},
  volume={10},
  number={1-3},
  pages={19--41},
  year={2000},
  publisher={Elsevier}
}

@phdthesis{shum2011unsupervised,
  title={Unsupervised methods for speaker diarization},
  author={Shum, Stephen},
  year={2011},
  school={Massachusetts Institute of Technology}
}

@article{campbell2006support,
  title={Support vector machines using GMM supervectors for speaker verification},
  author={Campbell, William M and Sturim, Douglas E and Reynolds, Douglas A},
  journal={IEEE signal processing letters},
  volume={13},
  number={5},
  pages={308--311},
  year={2006},
  publisher={IEEE}
}

@inproceedings{campbell2006svm,
  title={SVM based speaker verification using a GMM supervector kernel and NAP variability compensation},
  author={Campbell, William M and Sturim, Douglas E and Reynolds, Douglas A and Solomonoff, Alex},
  booktitle={Acoustics, Speech and Signal Processing, 2006. ICASSP 2006 Proceedings. 2006 IEEE International Conference on},
  volume={1},
  pages={I--I},
  year={2006},
  organization={IEEE}
}

@article{kenny2005joint,
  title={Joint factor analysis of speaker and session variability: Theory and algorithms},
  author={Kenny, Patrick},
  journal={CRIM, Montreal,(Report) CRIM-06/08-13},
  year={2005}
}

@article{dehak2011front,
  title={Front-end factor analysis for speaker verification},
  author={Dehak, Najim and Kenny, Patrick J and Dehak, R{\'e}da and Dumouchel, Pierre and Ouellet, Pierre},
  journal={IEEE Transactions on Audio, Speech, and Language Processing},
  volume={19},
  number={4},
  pages={788--798},
  year={2011},
  publisher={IEEE}
}

@inproceedings{lei2014novel,
  title={A novel scheme for speaker recognition using a phonetically-aware deep neural network},
  author={Lei, Yun and Scheffer, Nicolas and Ferrer, Luciana and McLaren, Mitchell},
  booktitle={Acoustics, Speech and Signal Processing (ICASSP), 2014 IEEE International Conference on},
  pages={1695--1699},
  year={2014},
  organization={IEEE}
}

@inproceedings{kenny2014deep,
  title={Deep neural networks for extracting baum-welch statistics for speaker recognition},
  author={Kenny, Patrick and Gupta, Vishwa and Stafylakis, Themos and Ouellet, P and Alam, J},
  booktitle={Proc. Odyssey},
  pages={293--298},
  year={2014}
}

@inproceedings{variani2014deep,
  title={Deep neural networks for small footprint text-dependent speaker verification},
  author={Variani, Ehsan and Lei, Xin and McDermott, Erik and Moreno, Ignacio Lopez and Gonzalez-Dominguez, Javier},
  booktitle={Acoustics, Speech and Signal Processing (ICASSP), 2014 IEEE International Conference on},
  pages={4052--4056},
  year={2014},
  organization={IEEE}
}

@inproceedings{chen2015multi,
  title={Multi-Task Learning for Text-Dependent Speaker Verification},
  author={Chen, Nanxin and Qian, Yanmin and Yu, Kai},
  booktitle={Sixteenth Annual Conference of the International Speech Communication Association},
  year={2015}
}

@inproceedings{povey2011kaldi,
  title={The Kaldi speech recognition toolkit},
  author={Povey, Daniel and Ghoshal, Arnab and Boulianne, Gilles and Burget, Lukas and Glembek, Ondrej and Goel, Nagendra and Hannemann, Mirko and Motlicek, Petr and Qian, Yanmin and Schwarz, Petr and others},
  booktitle={IEEE 2011 workshop on automatic speech recognition and understanding},
  number={EPFL-CONF-192584},
  year={2011},
  organization={IEEE Signal Processing Society}
}

@article{tieleman2012lecture,
  title={Lecture 6.5-rmsprop: Divide the gradient by a running average of its recent magnitude},
  author={Tieleman, Tijmen and Hinton, Geoffrey},
  journal={COURSERA: Neural networks for machine learning},
  volume={4},
  number={2},
  year={2012}
}

@article{richardson2015unified,
  title={A unified deep neural network for speaker and language recognition},
  author={Richardson, Fred and Reynolds, Douglas and Dehak, Najim},
  journal={arXiv preprint arXiv:1504.00923},
  year={2015}
}

@article{liu2015deep,
  title={Deep feature for text-dependent speaker verification},
  author={Liu, Yuan and Qian, Yanmin and Chen, Nanxin and Fu, Tianfan and Zhang, Ya and Yu, Kai},
  journal={Speech Communication},
  volume={73},
  pages={1--13},
  year={2015},
  publisher={Elsevier}
}

@inproceedings{fu2014tandem,
  title={Tandem deep features for text-dependent speaker verification.},
  author={Fu, Tianfan and Qian, Yanmin and Liu, Yuan and Yu, Kai},
  booktitle={INTERSPEECH},
  pages={1327--1331},
  year={2014}
}

@inproceedings{tian2015investigation,
  title={Investigation of bottleneck features and multilingual deep neural networks for speaker verification.},
  author={Tian, Yao and Cai, Meng and He, Liang and Liu, Jia},
  booktitle={Interspeech},
  pages={1151--1155},
  year={2015}
}

@inproceedings{li2015improved,
  title={Improved deep speaker feature learning for text-dependent speaker recognition},
  author={Li, Lantian and Lin, Yiye and Zhang, Zhiyong and Wang, Dong},
  booktitle={Signal and Information Processing Association Annual Summit and Conference (APSIPA), 2015 Asia-Pacific},
  pages={426--429},
  year={2015},
  organization={IEEE}
}

@inproceedings{heigold2016end,
  title={End-to-end text-dependent speaker verification},
  author={Heigold, Georg and Moreno, Ignacio and Bengio, Samy and Shazeer, Noam},
  booktitle={Acoustics, Speech and Signal Processing (ICASSP), 2016 IEEE International Conference on},
  pages={5115--5119},
  year={2016},
  organization={IEEE}
}

@article{hochreiter1997long,
  title={Long short-term memory},
  author={Hochreiter, Sepp and Schmidhuber, J{\"u}rgen},
  journal={Neural computation},
  volume={9},
  number={8},
  pages={1735--1780},
  year={1997},
  publisher={MIT Press}
}

@article{hinton2012deep,
  title={Deep neural networks for acoustic modeling in speech recognition: The shared views of four research groups},
  author={Hinton, Geoffrey and Deng, Li and Yu, Dong and Dahl, George E and Mohamed, Abdel-rahman and Jaitly, Navdeep and Senior, Andrew and Vanhoucke, Vincent and Nguyen, Patrick and Sainath, Tara N and others},
  journal={IEEE Signal Processing Magazine},
  volume={29},
  number={6},
  pages={82--97},
  year={2012},
  publisher={IEEE}
}

@article{larcher2014text,
  title={Text-dependent speaker verification: Classifiers, databases and RSR2015},
  author={Larcher, Anthony and Lee, Kong Aik and Ma, Bin and Li, Haizhou},
  journal={Speech Communication},
  volume={60},
  pages={56--77},
  year={2014},
  publisher={Elsevier}
}

@inproceedings{dey2016deep,
  title={Deep neural network based posteriors for text-dependent speaker verification},
  author={Dey, Subhadeep and Madikeri, Srikanth and Ferras, Marc and Motlicek, Petr},
  booktitle={Acoustics, Speech and Signal Processing (ICASSP), 2016 IEEE International Conference on},
  pages={5050--5054},
  year={2016},
  organization={IEEE}
}

@article{garofolo1993darpa,
  title={DARPA TIMIT acoustic-phonetic continous speech corpus CD-ROM. NIST speech disc 1-1.1},
  author={Garofolo, John S and Lamel, Lori F and Fisher, William M and Fiscus, Jonathon G and Pallett, David S},
  journal={NASA STI/Recon technical report n},
  volume={93},
  year={1993}
}

\end{document}